\documentclass[reprint,amsmath,amssymbGaps]{revtex4-1}

\usepackage{amsmath} 
\usepackage{amssymb}
\usepackage{amsfonts}
\usepackage{amstext}
\usepackage{graphicx}
\usepackage{bm}
\usepackage{color} 
\usepackage{transparent}
\usepackage{mathtools}
\usepackage{slashed}
\usepackage{hyperref}

\newcommand{\dal}{\Box \phi}
\newcommand{\dpp}{\left(\nabla \phi \right)^2}

\renewcommand{\a}{\alpha}
\renewcommand{\b}{\beta}

\begin{document}

\title{Derivation of Regularized Field Equations for the\\Einstein-Gauss-Bonnet Theory in Four Dimensions}

\author{Pedro G. S. Fernandes}
 \email{p.g.s.fernandes@qmul.ac.uk}
\author{Pedro Carrilho}
\author{Timothy Clifton}
\author{David J. Mulryne}
\affiliation{School of Physics and Astronomy, Queen Mary University of London, Mile End Road, London, E1 4NS, UK}


\begin{abstract}

\par We propose a regularization procedure for the novel Einstein-Gauss-Bonnet theory of gravity, which produces a set of field equations that can be written in closed form in four dimensions. Our method consists of introducing a counter term into the action, and does not rely on the embedding or compactification of any higher-dimensional spaces. This counterterm is sufficient to cancel the divergence in the action that would otherwise occur, and exactly reproduces the trace of the field equations of the original formulation of the theory. All other field equations display an extra scalar gravitational degree of freedom in the gravitational sector, in keeping with the requirements of Lovelock's theorem. We discuss issues concerning the equivalence between our new regularized theory and the original.

\end{abstract}

\maketitle

\section{Introduction}\label{intro}

Einstein's theory of General Relativity (GR) is the most successful theory of gravity we have, predicting and explaining a plethora of observations in the solar system and remote astrophysical systems \cite{Will:2014kxa}, as well as in the Universe at large \cite{Ishak:2018his}. These observations provide ample support for the validity of Einstein's theory, but at the same time it is also interesting to consider whether it is possible to construct alternative theories of gravity that might do equally well. Such possibilities are well studied in the physics literature, and can be motivated from high-energy physics \cite{Gross:1998jx} or the perceived shortcomings of GR \cite{ModifiedGravity1}.

In this regard, one of the most well studied classes of alternative theories of gravity are the Lovelock theories \cite{Lovelock_original}, which are specified by the following Lagrangian in $D$ dimensions:
\begin{equation}
\label{LoveL}
\mathcal{L} =  \sum_{j=0}^{n} \alpha_j \mathcal{R}^j \, ,
\end{equation}
where
\begin{equation}
\mathcal{R}^j = \frac{1}{2^j} \delta^{\mu_1 \nu_1 \dots \mu_j \nu_j}_{\alpha_1 \beta_1 \dots \alpha_j \beta_j} \prod_{i=1}^{j} R^{\alpha_i \beta_i}_{\phantom{\alpha_i \beta_i} \mu_i \nu_i} \, .
\end{equation}
The $\alpha_j$ in this equation are a set of $n$ arbitrary constants, $g$ is the determinant of the metric of the space-time, and $\displaystyle R^{\alpha_i \beta_i}_{\phantom{\alpha_i \beta_i} \mu_i \nu_i}$ are the components of the Riemann tensor. The symbol $\delta$ in the equation above is the generalized Kronecker delta, which is defined by
\begin{equation}
\delta^{\mu_1 \nu_1 \dots \mu_j \nu_j}_{\alpha_1 \beta_1 \dots \alpha_j \beta_j} \equiv j! \delta^{\mu_1}_{[\alpha_1} \delta^{\nu_1}_{\beta_1} \dots \delta^{\mu_j}_{\alpha_j} \delta^{\nu_j}_{\beta_j]} \, ,
\end{equation}
where the square brackets denote anti-symmetrization. The $n$ in Equation (\ref{LoveL}) can be taken to be given by $n=\frac{1}{2} (D-2)$ for even $D$, and $n=\frac{1}{2} (D-1)$ for odd $D$.

The Lovelock theories of gravity are of particular interest because they are the only Lagrangian-based theories of gravity that give covariant, conserved, second-order field equations in terms of the metric only. In this sense, they are the most natural possible generalizations of Einstein's theory. One way to see this is to write out the first few terms in Equation (\ref{LoveL}) explicitly, which gives
\begin{equation}
\label{Love3}
\mathcal{L} =  \left( - 2 \Lambda +R + \alpha_2 \mathcal{G} + \dots \right) \, ,
\end{equation}
where $\mathcal{G} \equiv R_{\alpha \beta \mu \nu} R^{\alpha \beta \mu \nu} - 4 R_{\mu \nu} R^{\mu \nu} +R^2$, and where we have chosen units such that $\alpha_1=1$ and defined $\alpha_0=-2\Lambda$. It can be seen that the first two terms in this equation correspond precisely to Einstein's theory with a cosmological constant, while the third term contains the quadratic Gauss-Bonnet term $\mathcal{G}$. All terms with higher-powers of the curvature tensors, which can be included in theories in dimensions $D \geqslant 6$, are included in the ellipsis.

In $D=5$ the Gauss-Bonnet term in Equation (\ref{Love3}) is well known to contribute to the field equations, and to produce a rich generalization of the phenomenology of Einstein's theory \cite{PhysRevLett.55.2656}. However, in $D=4$ the Gauss-Bonnet term has long been thought to have no have no consequences for the phenomenology of the classical theory. This is because the Chern theorem states that the contribution of $\mathcal{G}$ to the action one obtains from integrating the Lagrangian in Equation (\ref{Love3}) is entirely equivalent to a constant, proportional to the Euler characteristic of the space-time manifold \cite{10.2307/1969203}. It has therefore been thought to make no contribution to the field equations of the theory, which can be verified explicitly using dimensionally dependent identities for the curvature tensors \cite{Edgar:2001vv}.

Recently, however, a novel theory of gravity has been published that claims to bypass these difficulties \cite{original}. This theory has been dubbed {\it $4D$ Einstein-Gauss-Bonnet} (4DEGB) gravity, and is obtained by taking the coupling parameter $\alpha_2$ to scale as $1/(D-4)$ in the limit $D \rightarrow 4$. The idea here is to attempt to introduce a divergence that cancels the vanishing contribution that $\mathcal{G}$ makes to the field equations in four dimensions, in a manner that is conceptually similar to the dimensional regularization procedure used in quantum field theories. The goal of this is to produce a new classical gravity theory in four dimensions that includes a non-vanishing contribution from the Gauss-Bonnet term. It has attracted a great deal of recent attention  \cite{EGBST1, EGBST2, EGB2DIM, EGB1, EGB2, EGB3, EGB4, EGB5, EGB6, EGB7, EGB8, EGB9, EGB10, EGB11, EGB12, EGB13, EGB14, EGB15, EGB16, EGB17, EGB18, EGB19, EGB20, EGB21, EGB22, EGB23, EGB24, EGB25, EGB26, EGB27, EGB28, EGB29, EGB30, EGB31, EGB32}, 

In this work we investigate a method of regularizing the 4DEGB theory, in order to produce an action and set of field equations which are well defined in the limit $D \rightarrow 4$, and which can be written in closed form. Our method does not rely on the embedding or compactification of any higher-dimensional spaces, and results in a theory of gravity in which an extra scalar gravitational degree of freedom is made explicit. All solutions of the original 4DEGB theory published in \cite{original} are also found to be solutions of our new formulation of the theory.

The paper is organized as follows: In Section \ref{4degb} we summarize the original formulation of 4DEGB theory, as presented in \cite{original}. We then outline a regularization procedure for a divergent coupling constant used in two-dimensional gravity in Section\ref{2d}, following the procedure used in Ref. \cite{2Dpaper}. This procedure is then extended to the 4DEGB theory in Section \ref{GBtheory}. We discuss our results and conclude in Sections \ref{discussion} and \ref{conclusions}. Throughout, we use notation $\Box \equiv \nabla_\mu \nabla^\mu$ and $(\nabla \phi)^2 \equiv \nabla_\mu \phi \nabla^\mu \phi$, where $\nabla_{\mu}$ is the covariant derivative.

\section{4D Einstein-Gauss-Bonnet Gravity}
\label{4degb}

The novel 4DEGB theory was introduced in Ref.~\cite{original}, and is based on the action
\begin{equation}
\label{s4degb}
S[g_{\mu \nu}] = \int_{\mathcal{M}} d^Dx \sqrt{-g} \mathcal{L}_{\it EGB} +S_m \, ,
\end{equation}
where $S_m$ is the action associated with matter fields and $\mathcal{L}_{\it EGB}$ is the Einstein-Gauss-Bonnet Lagrangian given by
\begin{equation}
\mathcal{L}_{\it EGB} = - 2 \Lambda +R + \alpha \mathcal{G} \, ,
\end{equation}
which is the Lagrangian from Equation (\ref{Love3}) with all terms of cubic and higher powers in the curvature tensors neglected, and with $\alpha_2$ relabelled $\alpha$. The reader will note that the number of space-time dimensions $D$ is not yet specified.

The action (\ref{s4degb}) can then be varied with respect to the metric, and extremized, to yield the field equations
\begin{equation}
\label{fe}
G_{\mu \nu} + g_{\mu \nu} \Lambda = \alpha \mathcal{H}_{\mu \nu}+ T_{\mu \nu} \, ,
\end{equation}
where $T_{\mu \nu}$ is the energy-momentum tensor of matter and the contribution to the field equations from the Gauss-Bonnet term is given by
\begin{equation} \label{fetrace}
\mathcal{H}_{\mu \nu} = 15 \delta_{\mu [ \nu} R^{\rho \sigma}_{\phantom{\rho \sigma} \rho \sigma} R^{\alpha \beta}_{\phantom{\alpha \beta} \alpha \beta]} \, .
\end{equation}
The right-hand side of this equation is anti-symmetrized over five indices, and so must vanish in dimensions $D<5$.

Up until this point, the presentation has been the usual Einstein-Gauss-Bonnet theory. The novelty added in Reference \cite{original} is the possibility that the vanishing of $\mathcal{H}_{\mu\nu}$ in four dimensions might be cancelled by rescaling the coupling constant of the Gauss-Bonnet term such that $\alpha \rightarrow \hat{\alpha}/(D-4)$ in the limit $D\rightarrow 4$. That this might be a viable possibility is suggested by the trace of the field equations (\ref{fe}), which contain a contribution from the Gauss-Bonnet term that takes the form
\begin{equation}
\mathcal{H}^{\mu}_{\phantom{\mu} \mu} = \frac{1}{2} (D-4) \mathcal{G} \, ,
\end{equation}
where $\mathcal{G}$ is the Gauss-Bonnet term defined under Equation \eqref{Love3}. It is clear that in this case the multiplicative factor of $(D-4)$ would be precisely cancelled by the suggested rescaling of $\alpha$, which would leave a non-vanishing contribution to the trace of the field equations as $D \rightarrow 4$ (note that $\mathcal{G}$ itself is {\it not} required to vanish in this limit).

A proof that the off-diagonal field equations remain finite in the limit $D \rightarrow 4$ was not presented in Reference \cite{original}, but the authors of that paper did show that particular classes of solutions were well behaved in this limit. In particular, they showed that if they took $D$-dimensional Robertson-Walker geometries, with maximally symmetric spatial surfaces of dimension $D-1$, then the $D$-dimensional Friedmann equations were well behaved in the limit $D \rightarrow 4$. Similarly, the $D$-dimensional spherically symmetric vacuum solutions of the theory, with $(D-2)$-dimensionally spherically-symmetric subspaces, were also found to be well behaved in the appropriate limit. These results are suggestive that the theory may be well behaved in general in the four-dimensional limit.

The solutions to the 4DEGB theory, found using the prescription above, have some interesting features. The Friedmann equations, for example, contain corrective terms that are absent in the usual general relativistic equations, and the propagation equations for gravitational waves contain modified dispersion relations. Similarly, the new static and spherically symmetric black hole solutions differ from the Schwarzschild solution of GR, and exhibit a repulsive force as one approaches the central singularity. Remarkably, these new cosmological and black hole solutions formally coincide with the ones found in different contexts, \textit{e.g.} in gravity with a conformal anomaly and gravity with quantum corrections \cite{ConformalAnomaly1, ConformalAnomaly2, Quantum_Corrections2, Entropic_Corrections_FRW}.

Unfortunately, it now appears that there cannot be any closed-form expression for the full field equations of the novel Einstein-Gauss-Bonnet theory that is written in terms of curvature tensors only, and which remains finite as $D \rightarrow 4$ \cite{EGB31}. This is because the contribution of the Gauss-Bonnet term to the $D$-dimensional field equations can always be written as \cite{EGB31}
\begin{equation}
\mathcal{H}_{\mu \nu}= -2 (\mathcal{L}_{\mu \nu} + \mathcal{Z}_{\mu \nu}) \, ,
\end{equation}
where $\mathcal{L}_{\mu \nu}$ is written entirely in terms of the Weyl tensor as
\begin{equation}
\mathcal{L}_{\mu \nu} = C_{\mu \alpha \beta \sigma} C_{\nu}^{\phantom{\nu} \alpha \beta \sigma} - \frac{1}{4} g_{\mu \nu} C_{\alpha \beta \rho \sigma} C^{\alpha \beta \rho \sigma}
\end{equation}
and $\mathcal{Z}_{\mu \nu}$ contains all other contributions, such that
\begin{align}
\mathcal{Z}_{\mu \nu} =& \frac{(D-4) (D-3)}{(D-1)(D-2)} \Bigg[ -2 \frac{(D-1)}{(D-3)} C_{\mu \rho \nu \sigma} R^{\rho \sigma} \\ \nonumber &- 2 \frac{(D-1)}{(D-2)} R_{\mu \rho} R^{\rho}_{\phantom{\rho} \nu} + \frac{D}{(D-2)} R_{\mu \nu} R \\ \nonumber &+\frac{1}{(D-2)} g_{\mu \nu} \left( (D-1) R_{\rho \sigma} R^{\rho \sigma} - \frac{(D+2)}{4} R^2 \right) \Bigg] \, .
\end{align}
While it is clear that the contribution from $\mathcal{Z}_{\mu \nu}$ to the field equations will be regular after the rescaling $\alpha\rightarrow \hat{\alpha}/(D-4)$, no such statement can be made about $\mathcal{L}_{\mu \nu}$. This quantity vanishes identically in $D\leqslant 4$ only, and there does not appear to be any way to write it such that $\mathcal{L}_{\mu \nu} = (D-4) \mathcal{S}_{\mu \nu}$ for any well-behaved tensor $\mathcal{S}_{\mu \nu}$. The authors of Reference \cite{EGB31} further point out that simply allowing $\mathcal{L}_{\mu \nu}/(D-4)$ to vanish in the limit $D \rightarrow 4$ would not be sufficient to obtain well defined field equations, as in this case it would not be possible to satisfy the Bianchi identities. It therefore appears that there are no field equations for the novel 4D Einstein-Gauss-Bonnet theory that can be written in terms of curvature tensors only.

Without a well defined local action for the theory in the 4D limit, it is unclear how to properly count the dynamical degrees of freedom, how to establish whether the theory is covariant or well-posed, or how to find solutions in cases without explicit symmetries. In the next sections we will outline our remedy to this situation, and perform a regularization of the 4DEGB theory in order to produce a well defined, covariant set of field equations. This regularization makes explicit an extra scalar degree of freedom in the gravitational sector of the theory, and reproduces field equations that can also be obtained from a Kaluza-Klein dimensional reduction of $D>4$ Einstein-Gauss-Bonnet theory \cite{EGBST1, EGBST2}.

\section{Regularization in 2D}\label{2d}

The regularization procedure we wish to employ has already been successfully applied in two space-time dimensions, in order to construct an action for Einstein's equations, and we will use this section of our paper to outline its application in this case. We intend this to be an instructional demonstration of the methodology that will also be used in the four dimensional case in Section \ref{GBtheory}, to regularize the 4DEGB theory. This section closely follows the presentation used in Reference \cite{2Dpaper}.

Two-dimensional theories of gravity are known to be simpler than their four-dimensional counterparts, but nevertheless have been shown to admit rich and interesting structures (such as \textit{e.g.} black holes and cosmologies). Their simplicity also makes them a useful tool for the study of quantum gravity, which can be realised in this case \cite{1987MPLA....2..893P}. However, while it is possible to write down a consistent set of field equations, it is more problematic to write down an action from which the field equations can be derived. This is because in two dimensions the Einstein-Hilbert term has topological character, much like the Gauss-Bonnet term has in four dimensions, which has led to the construction of a number of gravitational theories (see, \textit{e.g.}, \cite{Strobl:1999wv} for a review on two-dimensional gravity). 

A recent development in two-dimensional theories of gravity, which is useful for our present study, is the development of a regularization procedure that introduces a divergence in the gravitational coupling parameter when $D\rightarrow 2$, and then cancels it out by adding a counter term to the action \cite{2Dpaper}. In this case we start by considering the following action in $D$ dimensions:
\begin{equation}
S= \alpha \int_{\mathcal{M}} d^D x \sqrt{-g} R + S_m \, ,
    \label{eq:EH}
\end{equation}
where $\alpha$ is a coupling constant. The contribution of $R$ to the field equations vanishes in $D=2$, so a divergence of the form $\alpha  \to   \hat{\alpha}/(D-2)$ is introduced into the coupling parameter in an attempt to get a non-zero result. This causes the action to become divergent, and of course highly reminiscent of the procedure introduced in the 4DEGB proposal.

In 2D, the construction of an equivalent theory is introduced by considering a conformally related geometry with metric
\begin{equation}
    \Tilde{g}_{\mu \nu} = e^{2\phi} g_{\mu \nu} \, , 
    \label{eq:conformal}
\end{equation}
where $\phi$ is a scalar function of the space-time coordinates. One can then add the following counter term to the action (\ref{eq:EH}):
\begin{equation}
    -\alpha \int_{\mathcal{M}} d^Dx \sqrt{-\Tilde{g}} \Tilde{R} \, ,
    \label{eq:assumption2d}
\end{equation}
where tildes denote the quantities constructed from $\Tilde{g}_{\mu \nu}$. 

One may note that in $D$ dimensions the square root of the determinant of the metric is related to its conformal counterpart by $\sqrt{-{g}}=e^{-D \phi}\sqrt{-\Tilde{g}}$, and that the Ricci scalar of the conformal metric can be specified as \cite{ConformalTransf1,ConformalTransf2}
\begin{align}
    \sqrt{-\Tilde{g}} \Tilde{R} = \sqrt{-g} e^{(D-2)\phi} \Big[& R - 2(D-1)\dal \\&- (D-1)(D-2)\dpp \Big]. \nonumber
\end{align}
Substituing this all into the action given by Equations (\ref{eq:EH}) and (\ref{eq:assumption2d}), and rescaling the coupling constant by $\alpha \rightarrow \hat{\alpha}/(D-2)$, one can find the result
\begin{align} \label{2d1}
	S=&\frac{\hat{\alpha}}{(D-2)} \int_{\mathcal{M}} d^D x \sqrt{-g} \Big[2(D-1)\dal \\ \nonumber &+ 2(D-1)(D-2)\dpp  -(D-2) \phi R \\ \nonumber &+ 2(D-2)(D-1)\phi \dal \Big] + S_m \, ,
\end{align}
where we have expanded the exponential around $D=2$ and discarded terms of order $\mathcal{O}((D-2)^2)$ or higher.

After performing an integration by parts on Equation (\ref{2d1}), we find that the factor $(D-2)$ cancels out precisely, and in the limit $D\rightarrow 2$ leaves
\begin{equation} \label{2daction}
    S=\hat{\alpha} \int_{\mathcal{M}} d^2 x \sqrt{-g} \left(\phi R + \dpp \right) + S_m \, ,
\end{equation}
where we stress that here $\hat{\alpha}$ is a finite constant in the limit $D \to 2$. This action has the field equations
\begin{equation} \label{2d3}
R = \frac{2}{\hat{\alpha}} T \, ,
\end{equation}
and
\begin{equation} \label{2d2}
\nabla_\mu \phi \nabla_\nu \phi - \nabla_\mu \nabla_\nu \phi + g_{\mu \nu} \left( \dal - \frac{1}{2} \dpp \right) = \frac{1}{\hat{\alpha}} T_{\mu \nu} \, ,
\end{equation}
where the stress-energy tensor obeys the conservation equation $\nabla^\mu T_{\mu \nu} = 0$, and has trace $T=T^{\mu}_{\phantom{\mu} \mu}$. The former of these two is the Einstein equation in two dimensions, while the trace of the latter gives $\dal = \frac{1}{\hat{\alpha}} T$. 

These field equations are particularly interesting as it can be seen that Equation (\ref{2d3}) can be equivalently written as $\Tilde{R}=0$, {\it i.e.} is equivalent to the vanishing of the Ricci curvature of the conformal geometry. This shows that the on-shell action of the regularized theory takes exactly the same form of the original theory (\ref{eq:EH}), and that the classical evolution of the gravity-matter system is independent of $\phi$ (although the converse is not true).

The theory with field equation (\ref{2d3}) is sometimes dubbed ``$R=T$" gravity, and has been studied in much detail in the literature \cite{TD1,TD2,TD3,TD4,TD5}. We remark that the regularized theory of \cite{2Dpaper} admits exactly the same solutions as $R=T$ gravity in $2D$ \cite{EGB2DIM,EGB28,2DBHsolution}, but that it also admits a finite and well-defined action. In the next section we will deploy a similar procedure to regularize the 4DEGB theory, in which we will obtain similar results.

\section{Regularization in 4D}\label{GBtheory}

The Einstein-Gauss-Bonnet theory in $D$ dimensions, and with vanishing cosmological constant, is described by the action
\begin{equation}
S=\int_{\mathcal{M}} d^D x \sqrt{-g} \left( R + \alpha \mathcal{G} \right) + S_m,
\label{EGBactionD}
\end{equation}
where $\alpha$ is a coupling constant. A cosmological constant can be trivially added later, and does not change any of the presentation that follows.

As discussed in Section \ref{4degb}, we want to consider this action in the presence of a coupling constant that is rescaled such that
\begin{equation}
\alpha \to \frac{\hat{\alpha}}{(D-4)} \, .
\label{eq:rescale}
\end{equation}
Following the procedure used in Section \ref{2d}, we consider a conformal geometry given by  $\Tilde{g}_{\mu \nu} = e^{2\phi} g_{\mu \nu}$, and add to the action a counterterm
\begin{equation}
    -{\alpha} \int_{\mathcal{M}} d^D x \sqrt{-\Tilde g} \Tilde{\mathcal{G}} \, ,
    \label{eq:assumption4d}
\end{equation}
where $\Tilde{\mathcal{G}}$ is the Gauss-Bonnet term constructed from the conformal metric $\Tilde{g}_{\mu \nu}$. We will find that this term removes the diverence that would otherwise occur in the action when the rescaling (\ref{eq:rescale}) is performed.

This can be seen by writing the Gauss-Bonnet term of the conformal metric in terms of the original one as
\begin{widetext}
\begin{equation}
    \begin{aligned}
\sqrt{-\Tilde g} \Tilde{\mathcal{G}} =&\sqrt{-g} e^{(D-4) \phi}\left[\mathcal{G}-8(D-3) R^{\mu \nu}\left(\nabla_{\mu} \phi \nabla_{\nu} \phi-\nabla_{\mu} \nabla_{\nu} \phi\right) \right. -2(D-3)(D-4) R\dpp \\&+4(D-2)(D-3)^{2}\dal\dpp -4(D-2)(D-3)\left(\nabla_{\mu} \nabla_{\nu} \phi\right)\left(\nabla^{\mu} \nabla^{\nu} \phi\right)+4(D-2)(D-3)\left(\dal\right)^2 \\
&+8(D-2)(D-3)\left(\nabla_{\mu} \phi \nabla_{\nu} \phi\right)\left(\nabla^{\mu} \nabla^{\nu} \phi\right)-4(D-3) R\dal \left.+(D-1)(D-2)(D-3)(D-4)(\nabla \phi)^4\right].
\end{aligned}
\end{equation}
Expanding the exponential around $D=4$, and neglecting terms of order $(D-4)^2$ or higher, we then obtain
\begin{equation}
\begin{aligned}
    \sqrt{-\Tilde g} \Tilde{\mathcal{G}} = \sqrt{-g} \Big( &\mathcal{G}-4(D-3)R\dal + 4(D-3)^2(D-2)\dal \dpp + 4(D-3)(D-2)(\dal)^2\\ 
    &- 8(D-3)R^{\mu \nu}(\nabla_\mu \phi \nabla_\nu \phi - \nabla_\mu \nabla_\nu \phi)+8(D-3)(D-2)\nabla_\mu \phi \nabla_\nu \phi \nabla^\mu \nabla^\nu \phi \\
    &- 4(D-3)(D-2)(\nabla_\mu \nabla_\nu \phi)(\nabla^\mu \nabla^\nu \phi)+\left(D-4\right) \Big[\phi \mathcal{G} -2(D-3)R\dpp \\&+(D-3)(D-2)(D-1)(\nabla \phi)^4 -4(D-3)\phi R \dal + 4(D-3)^2(D-2)\phi \dal \dpp\\&+4(D-3)(D-2)\phi (\dal)^2 - 8(D-3)\phi R^{\mu \nu}(\nabla_{\mu} \phi \nabla_{\nu} \phi-\nabla_{\mu} \nabla_{\nu} \phi) \\&+ 8(D-3)(D-2)\phi (\nabla_{\mu} \phi \nabla_{\nu} \phi)(\nabla^{\mu} \nabla^{\nu} \phi)- 4(D-3)(D-2)\phi (\nabla_{\mu} \nabla_{\nu} \phi)(\nabla^{\mu} \nabla^{\nu} \phi) \Big] \Big) \, .
\end{aligned}
\end{equation}
Integrations by parts, and making use of the identity
$$ \nabla_\mu \left[\Box \phi \nabla^\mu \phi - \frac{1}{2} \nabla^\mu \dpp \right] = \left( \Box \phi \right)^2  - \left(\nabla_\mu \nabla_\nu \phi\right)^2 - R^{\mu \nu} \nabla_\mu \phi \nabla_\nu \phi, $$
and the Bianchi identities, we can find that our action reads
\begin{equation}
\begin{aligned}
    S=\int_\mathcal{M} d^D x \sqrt{-g} \Big[ R + \alpha (D-4) \Big(&4(D-3)G^{\mu \nu}\nabla_\mu \phi \nabla_\nu \phi - \phi \mathcal{G} - 4(D-5)(D-3)\Box \phi \dpp \\& - (D-5)(D-3)(D-2) (\nabla \phi)^4 \Big) \Big] + S_m \, .
\end{aligned}
\end{equation}
After rescaling the coupling constant $\alpha$ as prescribed in \eqref{eq:rescale}, and taking the four-dimensional limit, this becomes
\begin{equation} \label{finals}
    S=\int_{\mathcal{M}} d^4 x \sqrt{-g} \Big[ R + \hat{\alpha} \Big(4G^{\mu \nu}\nabla_\mu \phi \nabla_\nu \phi - \phi \mathcal{G} + 4\Box \phi (\nabla \phi)^2 + 2(\nabla \phi)^4\Big) \Big] + S_m \, ,
\end{equation}
which can be seen to be free of divergences. This action belongs to the Horndeski class of theories \cite{Horndeski:1974wa,HorndeskiReview}, with functions $G_2=8 \hat{\alpha} X^2$, $G_3=8 \hat{\alpha} X$, $G_4=1+4 \hat{\alpha} X$ and $G_5 = 4 \hat{\alpha} \ln X$ (where $X=-\frac{1}{2} \nabla_{\mu} \phi \nabla^{\mu} \phi$). Note that the action is shift-symmetric in the scalar-field, yielding a conserved current whose divergence results in the scalar-field equation of motion \cite{ShiftSymmetry}.

The field equations of this new theory can be obtained by varying with respect to the metric, to get
\begin{equation}
    G_{\mu \nu} =  \hat{\alpha} \hat{\mathcal{H}}_{\mu \nu} +T_{\mu \nu} \, ,
    \label{eq:fieldeqs}
\end{equation}
where
\begin{equation}
\begin{aligned}
\hat{\mathcal{H}}_{\mu\nu} =&   2R(\nabla_\mu \nabla_\nu \phi - \nabla_\mu\phi \nabla_\nu \phi) + 2G_{\mu \nu}\left(\dpp-2\dal\right) + 4G_{\nu \a} \left(\nabla^\a \nabla_\mu \phi -\nabla^\a \phi \nabla_\mu \phi\right)\\
& + 4G_{\mu \a} \left(\nabla^\a \nabla_\nu \phi - \nabla^\a \phi \nabla_\nu \phi\right) + 4R_{\mu \a \nu \b}\left(\nabla^\b \nabla^\a \phi - \nabla^\a \phi \nabla^\b \phi\right)+ 4\nabla_\a\nabla_\nu \phi \left(\nabla^\a \phi \nabla_\mu \phi - \nabla^\a \nabla_\mu \phi \right)\\
& +4 \nabla_\a \nabla_\mu \phi \nabla^\a \phi \nabla_\nu \phi - 4\nabla_\mu \phi \nabla_\nu \phi \left(\dpp + \dal\right)+4\dal \nabla_\nu \nabla_\mu \phi - g_{\mu \nu} \Big( 2R\left(\dal - \dpp \right)\\
& + 4 G^{\a \b} \left( \nabla_\b \nabla_\a \phi - \nabla_\a \phi \nabla_\b \phi \right) + 2(\dal)^2 - \left( \nabla \phi\right)^4 + 2\nabla_\b \nabla_\a\phi\left(2\nabla^\a \phi \nabla^\b \phi - \nabla^\b \nabla^\a \phi \right)
\Big)\, ,
\end{aligned}
\end{equation}
and by varying with respect to the scalar field, to get
\begin{equation}
R^{\mu \nu} \nabla_{\mu} \phi \nabla_{\nu} \phi - G^{\mu \nu}\nabla_\mu \nabla_\nu \phi - \dal \dpp +(\nabla_\mu \nabla_\nu \phi)^2 - (\dal)^2 - 2\nabla_\mu \phi \nabla_\nu \phi \nabla^\mu \nabla^\nu \phi = \frac{1}{8}\mathcal{G} \, .
    \label{eq:gconf0}
\end{equation}
\end{widetext}

It is interesting to note that the trace of the field equations (\ref{eq:fieldeqs}) takes the simple form
\begin{equation} \label{4dtrace}
R+\frac{\hat{\alpha}}{2} \mathcal{G} = -T \, ,
\end{equation}
which is exactly the same form as the trace of the field equations of the original 4DEGB theory, as presented in Equations (\ref{fe}) and (\ref{fetrace}). Our theory therefore exactly reproduces the only known well defined field equation of the 4DEGB theory, and suggests that there may have been a hidden scalar degree of freedom in the original theory, which may be one reason it has not yet been proven possible to write its full field equations in terms of curvature tensors only (see Section \ref{4degb}).

Remarkably, the scalar field equation (\ref{eq:gconf0}) can be seen to be exactly equivalent to the condition
\begin{equation}
\Tilde{\mathcal{G}}=0 \, ,
\end{equation}
{\it i.e.} that the conformal Gauss-Bonnet term should vanish. This means that the counterterm we added to the action in Equation (\ref{eq:assumption4d}) must again vanish on shell, just as the corresponding term did in the 2D theory we discussed in Section \ref{2d}, and that our on-shell action has the same form as the action of the original 4DEGB theory. Note, however, that this fact does not guarantee that the theories are equivalent, but instead shows that solutions exist which solve both versions of the theory. We discuss the subject of equivalence further in the next section.

\section{Discussion}
\label{discussion}

We have used the regularization technique developed in Reference \cite{2Dpaper}, and applied it to the novel 4DEGB theory in order to find the regularized action (\ref{finals}). This action is free from divergences, and produces well behaved second-order field equations that can be used for gravitational physics. Our theory reproduces the trace of the field equations of the original theory (which is the only well defined field equation of the original 4DEGB theory), and complements it with a full set of off-diagonal equations.

We have been unable to show that the scalar degree of freedom $\phi$ decouples from the metric-matter system, except in the lone example of the trace equation, suggesting that the original theory may have a hidden scalar degree of freedom within it. If this is the case, then the 4DEGB theory does {\it not} propagate a single massless tensor degree of freedom, as claimed in the original paper \cite{original}. Instead, we find that the theory belongs to the Horndeski of scalar-tensor theories, and therefore does not bypass Lovelock's theorem. This hypothesis is backed-up by a recent study of the tree-level scattering amplitudes of gravitons in the original 4DEGB theory \cite{Bonifacio:2020vbk}.

We note that the action (\ref{finals}) is identical to the one that is obtained by performing a Kaluza-Klein reduction of a $(D+p)$-dimensional Einstein-Gauss-Bonnet theory with a flat $p$-dimensional internal space \cite{EGBST1, EGBST2}, as well as being the same action that appears in the context of renormalization group flows for trace anomalies of the effective action of the Nambu-Goldstone boson of broken conformal symmetry \cite{Komargodski:2011vj}. 

It is clear from Refs. \cite{EGBST1, EGBST2} that the cosmological and black hole solutions found in the original paper on novel 4DEGB gravity \cite{original} are also solutions of the field equations derived from the action (\ref{finals}). However, as discussed in these references, the action (\ref{finals}) also admits generalizations of these solutions. This includes a contribution to the Friedmann equations of the cosmological solutions that behaves like a fluid of radiation, or static black hole solutions with metric components $-g_{tt} \neq g_{rr}^{-1}$, which are absent in the solutions found in \cite{original}. 

While we believe our theory to be a compelling regularization of the original theory, we note that it is not possible to prove full equivalence of the two theories. This is because the original formulation of the theory does not have a full set of 4-dimensional field equations that can be written in closed form, but also because the dimensional regularization procedure used in \cite{original} does not appear to be unique. That is, there could potentially be arbitrarily many ways in which one could specify the geometry of the space-time before taking the limit $D \to 4$. There is no guarantee that all possibilities will yield the same solutions, and it is therefore very difficult to establish whether the set of admitted solutions of the two theories will always be the same.

The equivalence of the 2-dimensional theory presented in Reference \cite{2Dpaper}, and outlined in Section \ref{2d} does not suffer from the same difficulty. The $R=T$ theory and the field equations derived from the action (\ref{2daction}) are demonstrably identical, up to an additional equation that does not affect the metric-matter system. We expect this to be a feature of this procedure which is {\it only} applicable in 2-dimensions, as in this case there is only a single degree of freedom in the geometry, which means that the trace of the field equations contains all information about the theory. This is not true in dimensions $D>2$, so the equivalence of the trace equation (\ref{4dtrace}) does not directly imply equivalence of all of the field equations.

\section{Conclusions}
\label{conclusions}

We have investigated the application of the regularization procedure from Reference \cite{2Dpaper}, developed in the context of 2-dimensional gravity, to the novel 4DEGB theory recently proposed in Reference \cite{original}. We find that the counterterm we introduce in this procedure is sufficient to cancel the diverge in the action that would otherwise occur, and that the trace of the field equations (the only know field equation of the original formulation of the theory) is reproduced exactly. Our theory presents a full set of field equations that generalize this one equation to the full suite, and shows that an extra scalar field degree of freedom is also required in the gravitational sector of the theory.

The formulation of the theory that we end up with has an on-shell action that is identical to the action of the original theory, and produces second-order field equations that belong to the Horndeski class of scalar-tensor theories of gravity. The action is also identical to that which can be found from a Kaluza-Klein reduction of Einstein-Gauss-Bonnet theory in higher-dimensions \cite{EGBST1, EGBST2}, as well as  in the context of renormalization group flows \cite{Komargodski:2011vj}. It admits all of the solutions found in the original paper on novel 4DEGB theory, and provides a well defined set of equations that can be used to study the theory further.

\section*{Acknowledgements}

The authors would like to thank Tiago E. S. Fran\c{c}a and Jo\~{a}o F. Melo for useful discussions. PF is supported by the Royal Society grant RGF/EA/180022 and acknowledges support from the project CERN/FISPAR/0027/2019, PC and TC acknowledge financial support from the STFC under grant ST/P000592/1, and DJM is supported by a Royal Society University Research Fellowship.

\bibliography{biblio}

\end{document}